\documentclass[preprint,aps,floatfix,showpacs]{revtex4}       
\usepackage{graphicx}
\begin{document}

\title{Structures in high-energy fusion data}
\author{H. Esbensen}
\affiliation{$^1$Physics Division, Argonne National Laboratory, Argonne, 
Illinois 60439, USA,}
\date{\today}

\begin{abstract}
Structures observed in heavy-ion fusion cross sections at energies 
above the Coulomb barrier are interpreted as caused by the penetration 
of centrifugal barriers that are well-separated in energy.
The structures are most pronounced in the fusion of lighter, 
symmetric systems, where the separation in energy between successive 
angular momentum barriers is relatively large.
It is shown that the structures or peaks can be revealed by plotting 
the first derivative of the energy weighted cross section. 
It is also shown how an orbital angular momentum can be assign to 
the observed peaks by comparing to coupled-channels calculations.
This  is illustrated by analyzing high-energy fusion data for 
$^{12}$C+$^{16}$O and $^{16}$O+$^{16}$O, and the possibility of 
observing  similar structures in the fusion of heavier systems 
is discussed.
\end{abstract}
\maketitle

\section{Introduction}

The cross sections for the fusion of light, symmetric systems of nuclei 
sometimes exhibit structures or oscillations at energies above 
the Coulomb barrier. This has been observed both in measurements and
in coupled-channels calculations. The best experimental examples of 
this behavior are the fusion data of 
$^{12}$C+$^{12}$C \cite{sperrcc} $^{12}$C+$^{16}$O \cite{sperroc},
and $^{16}$O+$^{16}$O \cite{kovar,tserruya,kolata}.
The structures have been associated with resonances but there are also 
suggestions that they may be caused by the penetration of centrifugal 
barriers that are well-separated in energy \cite{vandenB,esbo16}. 

A simple reason the oscillations occur primarily in the fusion of 
lighter, symmetric systems is that the separation in energy between 
successive angular momentum barriers is relatively large in these systems.
Thus, when the separation of successive barriers becomes larger 
than twice the width associated with the penetration of the individual 
barriers, the oscillating pattern may occur.
This feature will be illustrated by applying a simple model that is based 
on the Hill-Wheeler barrier penetration formula \cite{hillwheel}. 

It was recently shown \cite{esbo16} that the structures observed 
in the $^{16}$O+$^{16}$O fusion data of Ref. \cite{tserruya} 
can be explained fairly well by coupled-channels calculations.
The calculations were based on a shallow potential in the entrance 
channel, whereas calculations based on a conventional Woods-Saxon
potential did not reproduce the data so well \cite{esbo16}. 
Thus there appears to be some connection between the oscillations 
in fusion cross sections at energies above the Coulomb barrier and 
the fusion hindrance phenomenon that occurs at deep subbarrier energies 
\cite{NiY,sys}. Both phenomena can be explained by applying a 
shallow M3Y+repulsion potential in coupled-channels calculations,
whereas a conventional Woods-Saxon potential fails \cite{esbo16,misi2}. 

A model of heavy-ion fusion which contains information about the 
centrifugal barriers at high energies is introduced in the  next 
section. It is based on the Hill-Wheeler formula for barrier
penetration \cite{hillwheel}, and it is shown how the first derivative
of the energy-weighted fusion cross section can be used as a diagnostic
tool to reveal the heights of the centrifugal barriers.
It is also shown why information about centrifugal barriers is lost 
in the commonly used Wong's formula \cite{wong}. 
In section III, the first derivative of the energy-weighted cross 
section is applied to analyze the structures that are observed in 
the fusion data of $^{12}$C+$^{12}$C, $^{16}$O+$^{16}$O, and 
$^{12}$C+$^{16}$O. The possibility of observing similar structures 
in the fusion of heavier systems is discussed in Section IV, and 
section V contains the conclusions.

\section{Model based on the Hill-Wheeler approximation}

In order to make a simple interpretation of high-energy fusion data 
one may resort to the well-known Hill-Wheeler formula \cite{hillwheel},
which expresses the barrier penetration probability in terms of a simple
Fermi function,
\begin{equation}
P_{\rm HW}(x) = \frac{\exp(x)}{1+\exp(x)},
\label{hwprob}
\end{equation}
where $x$ = $(E-V_{B}(L))/\epsilon_0$. Here $E$ is the center-of-mass 
energy, $V_B$ is the barrier height, and $\epsilon_0$ is a parameter 
that determines the exponential falloff at energies far below the barrier.
It can be derived from a parabolic approximation to the barrier potential 
but it is treated as an adjustable parameter in the following.  

The fusion cross section can now be obtained from the expression,
\begin{equation}
\sigma_f = \frac{\pi \ \hbar^2}{2\mu E} \sum_{L=0}^{L_{max}}
(2L+1) \ 
\frac{\exp(x_L)}{1+\exp(x_L)},
\label{sighw}
\end{equation}
where $x_L$ = $(E-V_B(L))/\epsilon_L$, and $\mu$ is the reduced mass
of the fusing system.
For a symmetric system of $0^+$ ground state nuclei, the sum over 
angular momenta in Eq. (\ref{sighw}) is restricted to even values of $L$ 
and the cross section must then be multiplied by a factor of 2. 
There are two types of parameters in Eq. (\ref{sighw}), namely, 
the $L$-dependent barrier heights, $V_B(L)$, and the decay constants 
$\epsilon_L$.

The first derivative of the energy-weighted cross section obtained from
Eq. (\ref{sighw}) is
\begin{equation}
\Bigl(\frac{d(E\sigma_f)}{dE}\Bigr)_{HW}  = 
\frac{\pi \ \hbar^2}{2\mu} \sum_{L=0}^{L_{max}}
(2L+1) \ \frac{1}{\epsilon_L} \ 
\frac{\exp(x_L)}{(1+\exp(x_L))^2}.
\label{hw1d}
\end{equation}
This expression can be interpreted as a sum of individual $L$-dependent 
barrier distributions weighted with the factor $(2L+1)$. 
Each distribution is centered at the barrier height $V_B(L)$, 
and it has a width that is determined by the constant $\epsilon_L$.
The constants $\epsilon_L$ are assumed in the following to be independent 
of $L$, i.~e., $\epsilon_L$ = $\epsilon_0$.

\subsection{Wong's formula}

The model (\ref{sighw}) was applied by Wong \cite{wong} to derive his 
formula for the fusion cross section. He assumed that the $L$-dependent
barriers were parametrized as follows,
\begin{equation}
V_B(L) = V_{CB} + 
\frac{\hbar^2 L(L+1)}{2\mu R_{CB}^2},
\label{centro}
\end{equation} 
where $V_{CB}$ is the height of the Coulomb barrier (for $L$=0),
$\mu$ is the reduced mass of the system, and $R_{CB}$ is
the radial distance at the Coulomb barrier. 
By replacing the discrete sum over $L$ in Eq. (\ref{sighw})
with a continuous integration over $L$, 
i.~e., $\sum_L(2L+1)$ $\rightarrow$ $\int d[L(L+1)]$,
Wong obtained the following compact formula \cite{wong},
\begin{equation}
\sigma_f = \pi R_{CB}^2 \ \frac{\epsilon_0}{E} \
\ln(1 + \exp(x_0)),
\label{wong}
\end{equation}
where $x_0$ = $(E-V_{CB})/\epsilon_0$.
The first derivative of Wong's formula,
\begin{equation}
\Bigl(\frac{d(E\sigma_f)}{dE}\Bigr)_W = 
\pi R_{CB}^2 \ \frac{\exp(x_0)}{1+\exp(x_0)},
\label{wong1d}
\end{equation}
is proportional to a Fermi function and it approaches the value
$\pi R_{CB}^2$ at energies far above the Coulomb barrier. 

The barrier distribution for heavy-ion fusion reactions that was 
introduced in Ref. \cite{rowley} is defined as the second derivative 
of the energy weighted cross section,
\begin{equation}
B(E) = \frac{d^2(E\sigma_f)}{dE^2}. 
\label{barr}
\end{equation}
This definition was partly inspired by Wong's formula because the
second derivative one obtains in this case,
\begin{equation}
\Bigl(\frac{d^2(E\sigma_f)}{dE^2}\Bigr)_W = 
\pi R_{CB}^2 \ \frac{1}{\epsilon_0} \ 
\frac{\exp(x_0)}{\bigl[1+\exp(x_0)\bigr]^2},
\label{wong2d}
\end{equation}
is a nice symmetric distribution that is centered at the Coulomb barrier 
$V_{CB}$ (for $L$=0.) The width is determined by $\epsilon_0$, which 
characterizes the exponential falloff of barrier penetrability at 
energies far below the $s$-wave barrier.

The definition Eq. (\ref{barr}) is reasonable at energies close to 
the Coulomb barrier. However, it does not reveal any information about
the individual $L$-dependent barriers. A better way to search for 
evidence of the individual centrifugal barriers is to plot the 
first derivative of the energy weighted cross sections, according 
to the Hill-Wheeler expression, Eq. (\ref{hw1d}).

In order to be able to identify the individual centrifugal barriers 
from the measured fusion cross sections, it is necessary that the 
energy difference between successive barriers is much larger than 
twice the width of the individual barriers. Using the simple 
expression, Eq. (\ref{centro}), 
one obtains the following expression for the energy 
difference between the heights of successive barriers, 
\begin{equation}
\Delta V_B = V_B(L+1) - V_B(L)
\approx 
\frac{\hbar^2 2(L+1)}{2\mu R_{CB}^2}.
\label{deb}
\end{equation} 
The width of the individual barrier distributions that appear in 
Eq. (\ref{hw1d}) is characterized by the parameter $\epsilon_L$, 
which is assumed to be independent of $L$ and equal to $\epsilon_0$. 
The requirement that the energy difference, Eq. (\ref{deb}), 
is much larger than 2$\epsilon_0$ implies that 
\begin{equation}
(L+1) \ {\rm >>} \ \frac{2\mu R_{CB}^2\epsilon_0}{\hbar^2}.
\label{condi1}
\end{equation} 
This condition applies to the fusion of an asymmetric system, 
where the fusion can occur for all values of $L$.
For a symmetric system of $0^+$ ground state nuclei, the fusion
can only take place for even values of the angular momentum.
The condition for observing individual barriers 
is then replaced by
\begin{equation}
(2L+3) \ {\rm >>} \ \frac{2\mu R_{CB}^2\epsilon_0}{\hbar^2}.
\label{condi2}
\end{equation} 

\subsection{Illustration of Hill-Wheeler's formula}

\begin{figure}
\includegraphics[width = 8cm]{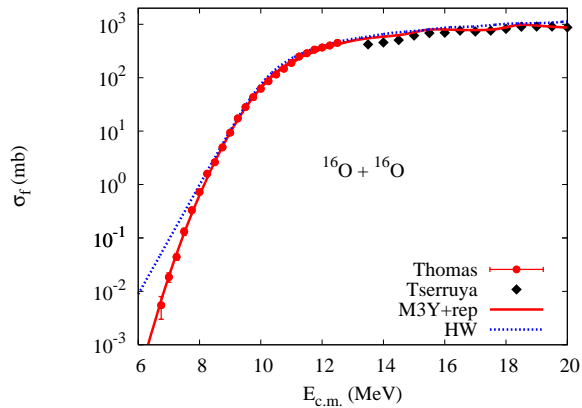}
\caption{\label{fuso16} 
(Color online) 
Measured cross sections for the fusion of $^{16}$O+$^{16}$O \cite{thomas,tserruya} 
are compared to the simplified Hill-Wheeler model (HW) with the parameters 
$V_{CB}$ = 9.9 MeV, $R_{CB}$=8.4 fm, $\epsilon_0$ = 0.4 MeV.
The solid curve (M3Y+rep) is the result of the coupled-channels calculation 
presented in Ref.  \cite{esbo16}.}
\end{figure}

Measured cross sections for the fusion of $^{16}$O+$^{16}$O 
\cite{thomas,tserruya} are compared in Fig. \ref{fuso16}
to the Hill-Wheeler cross section (HW), Eq. (\ref{sighw}), using the 
parameters $V_{CB}$ = 9.9 MeV, $\epsilon_0$ = 0.4 MeV and $R_{CB}$=8.4 fm. 
These parameters provide a fair representation of the 
data above 8 MeV and are used below for illustrative purposes.
Also shown is the coupled-channels calculation of Ref. 
\cite{esbo16} (solid curve) that is discussed in
more detail in the next sections.

\begin{figure}
\includegraphics[width = 8cm]{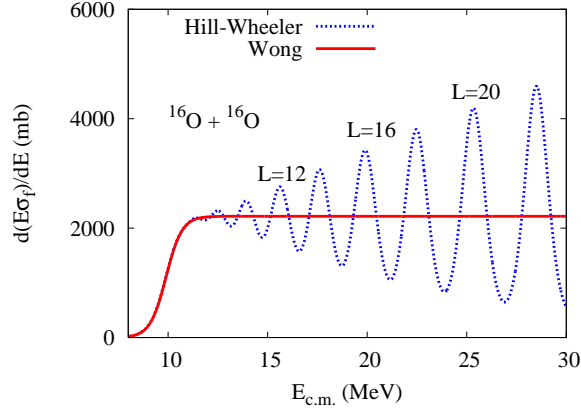}
\caption{\label{f1dhwo16} 
(Color online) 
First derivative of the energy-weighted fusion cross sections 
for $^{16}$O+$^{16}$O obtained from Hill-Wheeler's and Wong's formulas.
The parameters are the same as in Fig. \ref{fuso16}.
The expression obtained from Wong's formula, Eq. (\ref{wong1d}), behaves 
like a Fermi function and approaches $\pi R_{CB}^2$ at high energy. The 
Hill-Wheeler expression, Eq. (\ref{hw1d}), has the same behavior in the 
vicinity of the Coulomb barrier but starts to oscillate at high energies. 
The peaks show the location of the individual centrifugal barriers;
the barriers for $L$ = 12, 16 and 20 are indicated.}
\end{figure}

The first derivative of the energy-weighted cross section for the 
fusion of $^{16}$O+$^{16}$O is illustrated in Fig.  \ref{f1dhwo16}.
in terms of Wong's and Hill-Wheeler's formulas.
The parameters used here are the same as those mentioned above.
The first derivative of Wong's formula is a Fermi function which 
approaches the constant value $\pi R_{CB}^2$ at energies far above 
the Coulomb barrier, c.~f. Eq. (\ref{wong1d}). 
The Hill-Wheeler expression, Eq. (\ref{hw1d}), reproduces this
behavior at energies near and below the Coulomb barrier but it 
starts to oscillate at energies above the Coulomb barrier.
The peaks in this figure reflect the location of the individual 
centrifugal barriers, as evidenced by Eq. (\ref{hw1d}). 
The peaks for $L$ = 12, 16, and 20 are labeled in the figure.

The lowest peak that is visible in Fig. \ref{wong1d} is due to the 
centrifugal barrier for $L$ = 8. This observation is consistent with 
the condition, Eq. (\ref{condi2}), which in the example considered 
here requires that $(2L+3)$ $>>$ 11. The condition is not fulfilled 
for $L$ = 4 or 6, but it is reasonably well satisfied for $L$ = 8.
As the angular momentum increases, the overlap between neighboring
peaks diminishes which results in the breakdown of Wong's formula.
The breakdown of Wong's formula is primarily a problem in light,
symmetric systems, whereas it is usually not recognized in the 
fusion of heavy systems.

\begin{figure}
\includegraphics[width = 8cm]{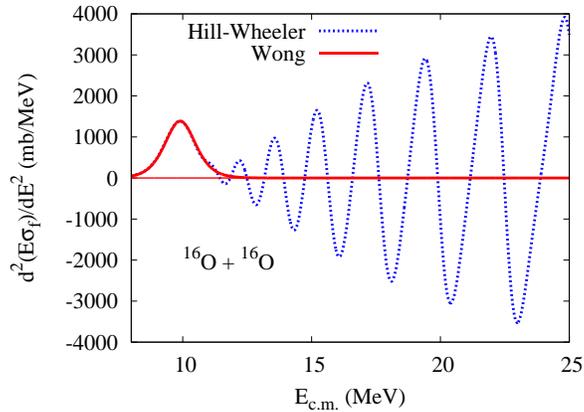}
\caption{\label{f2dhwo16} 
(Color online) 
Second derivative of the energy-weighted fusion cross sections
for $^{16}$O+$^{16}$O obtained from Hill-Wheeler's formula, Eq. (\ref{sighw}), 
and from Wong's formula, Eq. (\ref{wong}). The parameters for the two 
expressions are quoted in the caption of Fig. 1.
They produce almost identical results in the vicinity of the Coulomb barrier 
($V_{CB}$=9.9 MeV). The second derivative of Wong's formula goes to zero at 
high energies, whereas the Hill-Wheeler expression starts to oscillate.}
\end{figure}

Let us for completeness also examine the conventional barrier 
distribution \cite{rowley}, i.~e., the second derivative of the 
energy weighted cross section. The results one obtains in the 
example considered in this subsection are shown in Fig. \ref{f2dhwo16}.
Wong's formula gives a symmetric distribution, Eq. (\ref{wong2d}),
which is peaked at the Coulomb barrier.
The distribution derived from Hill-Wheeler's formula reproduces 
Wong's formula in the vicinity of the Coulomb barrier but it starts 
to oscillate at higher energies. This illustrates vividly the 
breakdown of Wong's formula. It should be emphasized that the peaks 
in Fig. \ref{f2dhwo16} at high energies do not represent the actual 
centrifugal barrier distributions; the barrier distributions at 
high energy are depicted in Fig. \ref{f1dhwo16}. It is only the 
peak at the lowest energy in Fig. \ref{f2dhwo16} that represents 
a real barrier distribution, and it is associated with the angular
momentum $L$ = 0.

\section{Structures in light-ion fusion}

In this section the high-energy fusion data for $^{12}$C+$^{12}$C \cite{sperrcc}, 
$^{12}$C+$^{16}$O \cite{sperroc}, and $^{16}$O+$^{16}$O \cite{tserruya} are 
compared to simple estimates based on the Hill-Wheeler formula and to 
coupled-channels calculations.
The comparison is made in terms of the first derivative of the energy
weighted cross section, which in the following is defined in terms of 
the average, finite difference value,
\begin{equation}
(\frac{d(E\sigma)}{dE})_n = 
\frac{1}{2} \Bigl[
\frac{(E\sigma)_{n+1}-(E\sigma)_n}{E_{n+1}-E_{n}} \ + \
\frac{(E\sigma)_{n}-(E\sigma)_{n-1}}{E_{n}-E_{n-1}}\Bigr].
\label{disc}
\end{equation}
This definition is used to determine both the calculated and measured values.
The energies $E_n$ are the discrete energies where the measurements/calculations
are performed. The average energy associated with the definition (\ref{disc})
is ${\bar E}_n=(E_{n-1}+2E_n+E_{n+1})/4$.

The coupled-channels calculations are solved in the rotating frame
approximation with ingoing-wave boundary conditions (IWBC) that are 
imposed at the minimum of the pocked of the entrance channel potential
\cite{misi2}.
The fusion cross section is obtained from the ingoing flux at the boundary.
A slight improvement is to impose the IWBC for each orbital angular
momentum $L$ at the minimum of the pocket in each centrifugal potential.
This definition works quite well at energies near and below the
Coulomb barrier but it can be difficult to account for the data at
high energy. The problem can be solved by introducing an imaginary 
potential \cite{misi2},
\begin{equation}
W(r) = \frac{W_0}{1+\exp((r-R_w)/a_w)},
\end{equation}
where the radius parameter $R_w$ is chosen to coincide with the
location of the pocket minimum. At low energies, it is sufficient 
to use a weak and short ranged potential, with typical parameters
$W_0$ = -2 MeV and $a_w$ = 0.2 fm.
 
At high energies it is often necessary to increase the values of the 
parameters $W_0$ and $a_w$ if one wants to account for the data. 
This problem seems to be more serious for heavy systems where a large 
number of reaction channels that are not treated explicitly in the 
coupled-channels calculations open up. There are other aspects of the 
calculations that are questionable as the maximum angular momentum 
for fusion is reached. For example, angular momentum dissipation
must play an important role at very high energies and the rotating 
frame approximation must therefore become questionable. These
issues will not be addressed here.


\subsection{Fusion of $^{16}$O+$^{16}$O} 

\begin{figure}
\includegraphics[width = 7cm]{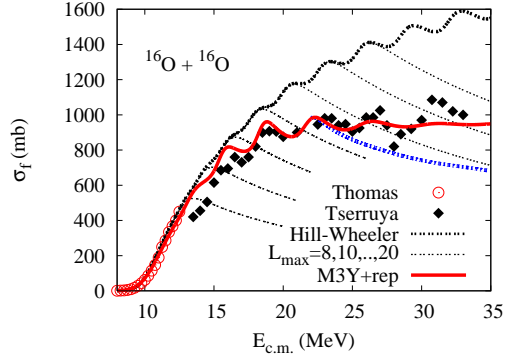}
\caption{\label{flfo16cc} 
(Color online) The measured fusion cross sections for $^{16}$O+$^{16}$O 
\cite{thomas,tserruya} are compared to the coupled-channels calculations
(solid red curve) \cite{esbo16} that are based on the M3Y+repulsion potential.
The thick (blue) dashed curve is the coupled-channels result for a maximum 
angular of $L_{max}$ = 16.
The thick, black dashed curve is the prediction of Hill-Wheeler's formula,
whereas the thin dashed curves are the 
predictions for $L_{max}$ = 8, 10,...,20.}
\end{figure}

The cross section one obtains by applying the Hill-Wheeler parametrization 
(\ref{sighw}) to the fusion of $^{16}$O+$^{16}$O is compared in Fig. 
\ref{flfo16cc} to the high-energy data of Tserruya et al. \cite{tserruya}
and also to the low-energy data of Thomas et al. \cite{thomas}.
The parameters are the same as used in the previous section.
These parameters provide a fairly good description of the data at 
energies near the Coulomb barrier (and above 8 MeV) as illustrated 
in Fig. 1, but they fail to account for the high-energy data. 
One interpretation is that the centrifugal barriers predicted by Eq.  
(\ref{centro}) and the parameters considered here are not correct at 
high angular momenta.

The Hill-Wheeler cross sections one obtains for different choices of
the maximum angular momentum for fusion, namely, $L_{max}$ = 8, 10, ..., 20, 
are also shown in Fig. \ref{flfo16cc}. They fall off as $1/E$ when 
the energy exceeds the height 
of the maximum barrier considered. 
To be specific, the cross section for a symmetric system behaves like
\begin{equation}
\sigma_f(E,L_{max}) \approx \frac{\pi\hbar^2}{2\mu E} \ 
[L_{max}(L_{max}+3)+2],
\label{sigcut}
\end{equation}
when $E >> V_B(L_{max})$. A similar expression holds for asymmetric
systems with the $L_{max}$ dependent factor replaced by
$(L_{max}+1)^2$.  The simple dependence on energy and  
maximum angular momentum is very useful because it can be used 
to roughly assign an angular momentum associated to each
centrifugal barrier extracted from an experiment. 

The fusion data shown in Fig. \ref{flfo16cc} follow the predicted 
$1/E$ dependence in small sections of energy but they do not agree 
with the magnitude of the curves predicted for different values of 
$L_{max}$. 
In fact, the data fall mostly halfway 
between these curves when the $1/E$ dependence occurs. 
This is a somewhat disturbing feature but it is nicely reproduced 
by the coupled-channels calculation of Ref.  \cite{esbo16}, which 
is shown by the solid (red) curve. The blue dashed curve shows 
the coupled-channels result one obtains by imposing a maximum 
angular momentum of $L_{max}$ = 16. This curve does eventually 
approach the the Hill-Wheeler prediction for $L_{max}$ =16 but 
it occurs at an almost 10 MeV higher energy. 

The coupled-channels calculation of Ref. \cite{esbo16} was calibrated 
to reproduce the low-energy fusion data of Thomas et al. \cite{thomas}. 
It was supplemented with a short-ranged imaginary potential that
acts near the minimum of the pocket in the entrance channel 
potential in order to improve the behavior at high energy.
In spite of the latter adjustment, it is remarkable that the  
calculation does reproduce the high energy data so well
up to about 27 MeV.

\begin{figure}
\includegraphics[width = 7cm]{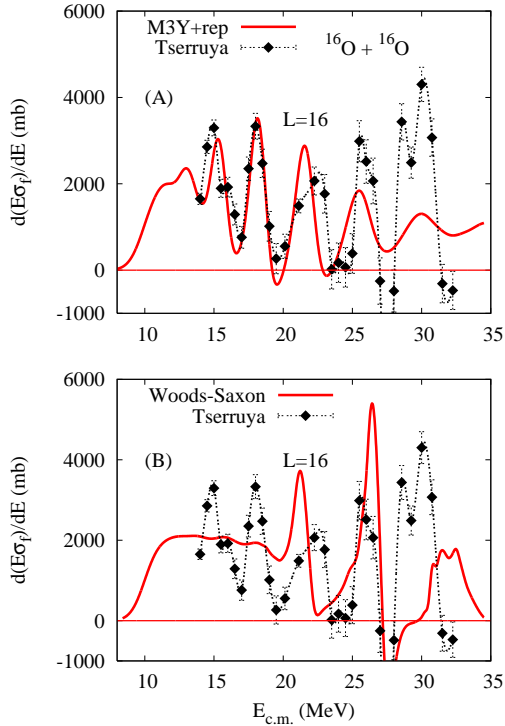}
\caption{\label{f1do16cc} 
(Color online)
First derivative of the measured energy-weighted cross section for 
the fusion of $^{16}$O+$^{16}$O \cite{tserruya} is compared to 
coupled-channels calculations \cite{esbo16} that are based on the 
M3Y+repulsion potential (A), and on a conventional Woods-Saxon potential (B).
The calculated peaks at $L$ = 16 are marked.}
\end{figure}

A good way to amplify the structures in the high energy data is 
to plot the first derivative of the energy weighted cross section.  
The result are shown in Fig. \ref{f1do16cc}A. The structures of 
the data are reproduced remarkably well in this representation 
by the coupled-channels calculation (the thick solid curve.)
That gives confidence in the assignment of an angular momentum 
to each individual peak because the angular momenta of the 
calculated peaks are well determined. 
The peak associated with $L$ = 16 is marked in the figure for
clarity so that the peaks for $L$ = 12 to 20 can easily be identified. 

A similar coupled-channels calculation, which was based on a 
conventional Woods-Saxon potential, was also performed in Ref. 
\cite{esbo16}. It did a rather poor job in reproducing the data 
at high energy (see Fig. 7 of Ref.  \cite{esbo16}.) 
The barrier distributions one obtains from this calculation are 
shown in Fig. \ref{f1do16cc}B.
While the location of the $L$=16 peak is essentially the same as 
obtained with the M3Y+repulsion potential, the structures at smaller 
angular momenta have essentially disappeared. By comparing the two 
figures, Figs. \ref{f1do16cc}A and \ref{f1do16cc}B, it is clear that 
the M3Y+repulsion potential provides the better description of the data.


\subsection{Fusion of $^{12}$C+$^{12}$C} 

Another example of a system that exhibits strong structures in 
its high-energy fusion data is $^{12}$C+$^{12}$C  
\cite{sperrcc}. Unfortunately, it was not possible to reproduce 
the data so well by coupled-channels calculations, as it was done
for the fusion of $^{16}$O+$^{16}$O. It is therefore of interest to 
try a different approach when analyzing the data. One way 
is to use the Hill-Wheeler parametrization of 
the cross section, Eq. (\ref{sighw}), and treat the energies of the 
centrifugal barriers, $V_{B}(L)$, as adjustable parameters.

\begin{figure}
\includegraphics[width = 12cm]{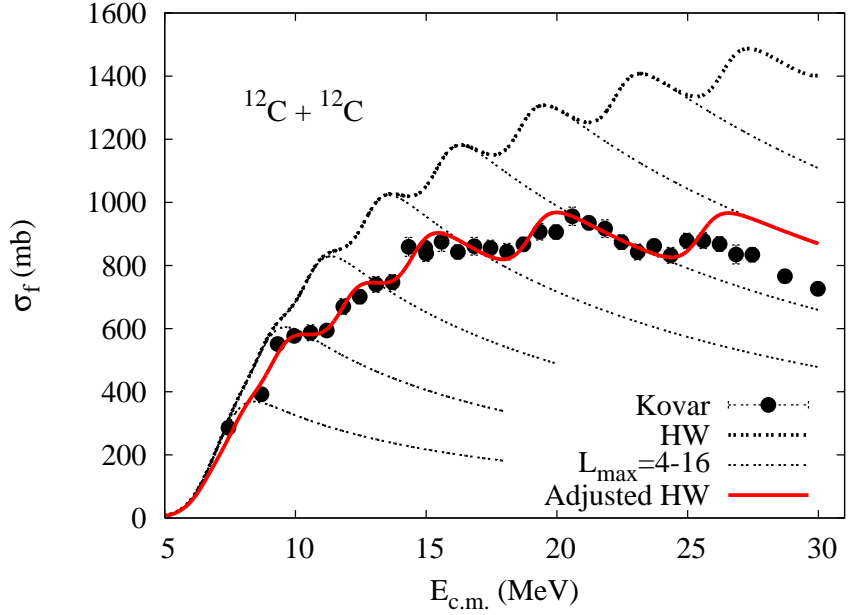}
\caption{\label{flfc12hw} 
(Color online)
Cross sections for the fusion of $^{12}$C+$^{12}$C \cite{sperrcc}
are compared to the Hill-Wheeler expression (HW) using the parameters
$V_{CB}$ = 6.23 MeV, $R_{CB}$=7.667 fm, and $\epsilon_0$ = 0.4 MeV.
The solid curve was obtained by adjusting the heights of the 
centrifugal barriers to optimize the fit to the data below 25 MeV.} 
\end{figure}

The high-energy fusion data for $^{12}$C+$^{12}$C that were measured 
Sperr et al. \cite{sperrcc} are shown in Fig. \ref{flfc12hw}.
Also shown is a prediction by Hill-Wheeler's formula, Eq. (\ref{sighw}), 
which is based on the parameters:
$V_{CB}$ = 6.23 MeV, $\epsilon_0$ = 0.4 MeV and $R_{CB}$= 7.667 fm. 
These parameters characterizes the entrance channel potential that 
was used in the coupled-channels calculations of Ref. \cite{esbc12}.
The cross sections for different maximum angular momentum cutoffs, namely,
for $L_{max}$ = 4-16, are also shown. They coincide in most cases with the 
data when the data exhibit the characteristic $1/E$ dependence.
One exception is at the highest energies, above 25 MeV, where the data 
fall half-way between the predictions for $L_{max}$ = 12 and 14.

\begin{figure}
\includegraphics[width = 12cm]{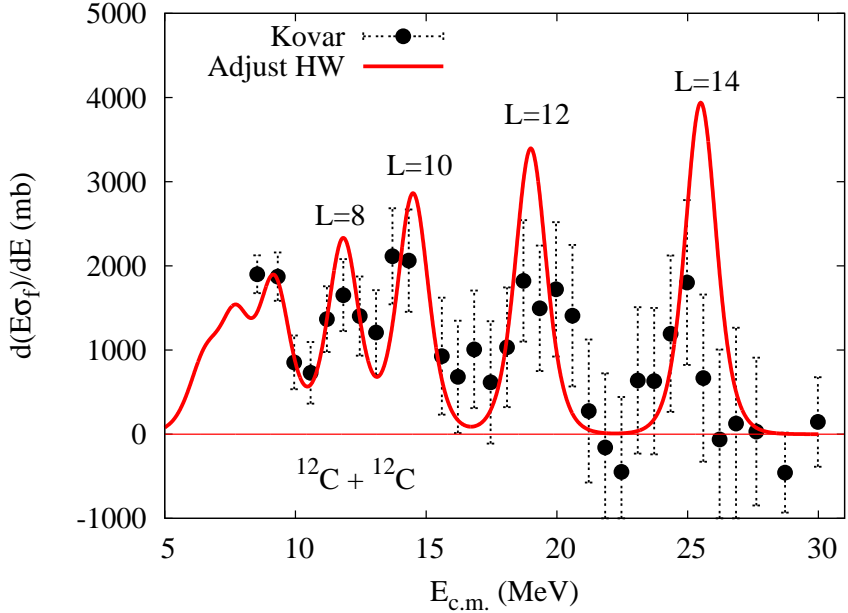}
\caption{\label{f1dc12hw} 
(Color online)
First derivative of the energy-weighted cross sections shown in 
Fig. \ref{flfc12hw}.}
\end{figure}

The fact that the data agree so well with the $1/E$ curves 
when the data exhibit the characteristic $1/E$ behavior,
makes it fairly easy to fit the data simply by adjusting the heights 
of the centrifugal barriers.
The result that gives the best fit to the data up to 25 MeV is shown 
by the solid curve in Fig. \ref{flfc12hw}. A more detailed comparison 
is shown in Fig. \ref{f1dc12hw} in terms of the first derivative of the 
energy-weighted cross section. Here one can see that the widths of the 
measured centrifugal barrier distributions apparently increase with 
increasing angular momentum, whereas the width was assumed to be a
constant characterized by the parameter $\epsilon_0$ in the calculation. 
It may be useful to incorporate an $L$-dependence of the value of
$\epsilon_L$ in the data analysis but that will not be tried here. 
The ultimate goal is to develop a coupled-channels calculation 
that can account for the data
and provide a more reliable determination of the angular momenta
associated with the experimental peaks shown in Fig. \ref{f1dc12hw}.  

\subsection{Fusion of $^{12}$C+$^{16}$O} 

\begin{figure}
\includegraphics[width = 12cm]{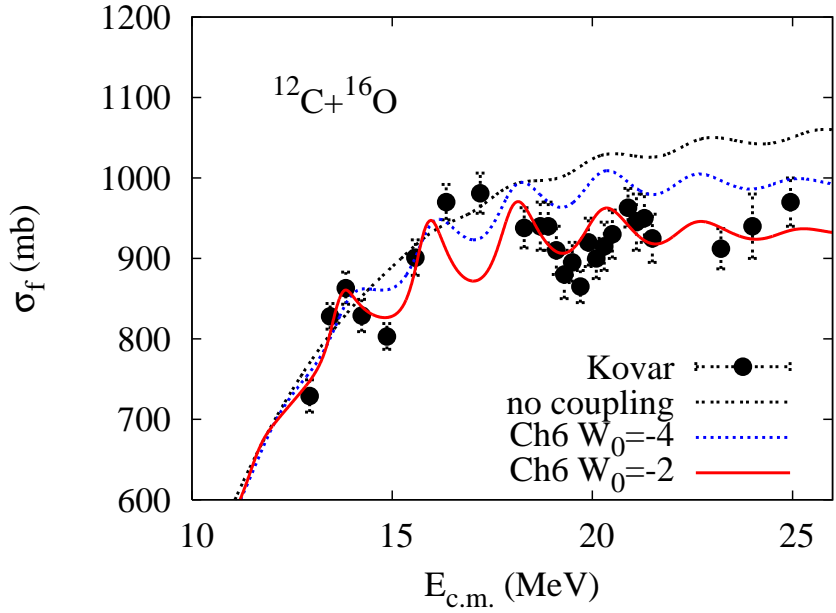}
\caption{\label{flfc12o16} 
(Color online)
Cross sections for the fusion of $^{12}$C+$^{16}$O \cite{sperroc} are 
compared to coupled-channels calculations (Ch6) with different strengths 
of the imaginary potential ($W_0$= -2 and -4 MeV, respectively), 
and to the no-coupling limit (top dashed curve).}
\end{figure}

The last example in this section is the fusion of $^{12}$C+$^{16}$O
which was also measured by Sperr et al. \cite{sperroc}. This is an 
asymmetric system so the fusion can occur for all values of $L$.
The data are compared in Fig. \ref{flfc12o16} to coupled-channels 
calculations that include 6 channels (Ch6).
The six channels are the elastic channel, the four channels associated
with the excitation of the $2^+$ and $3^-$ states in either projectile
or target, and the channel associated with the excitation of the
$0_2^+$ in $^{12}$C. The structure input to the calculation can be 
found in the papers on the fusion of oxygen plus oxygen \cite{esbo16}
and carbon plus carbon \cite{esbc12}.

The M3Y+repulsion, double-folding potential that is used in the 
calculation is generated from the charge densities of the reacting nuclei.
The repulsive part of the interaction is determined by the
incompressibility $K$ = 234 MeV, and the diffuseness parameter $a_r$
associated with the repulsion \cite{misi2}.
The latter was set to $a_r$ = 0.41 fm because that was the preferred 
value in the analysis \cite{esbo16} of the $^{16}$O+$^{16}$O fusion 
data by Thomas et al. \cite{thomas}.

The sensitivity to the imaginary potential is illustrated in 
Fig. \ref{flfc12o16} by two coupled-channels calculations, one
with the strength $W_0$ = -2 MeV and one with $W_0$ = -4 MeV,
whereas the diffuseness was kept fixed at $a_w$ = 0.2 fm.
It appears that the data are best described by the weaker absorption. 
The top dashed curve is the result of the no-coupling calculation 
which employs the weak absorption, $W_0$ = -2 MeV. Here the 
oscillations in the high energy cross sections are modest.
Evidently, the strong structures in the solid curve of Fig. 
\ref{flfc12o16} are caused by coupled-channels effects.

\begin{figure}
\includegraphics[width = 12cm]{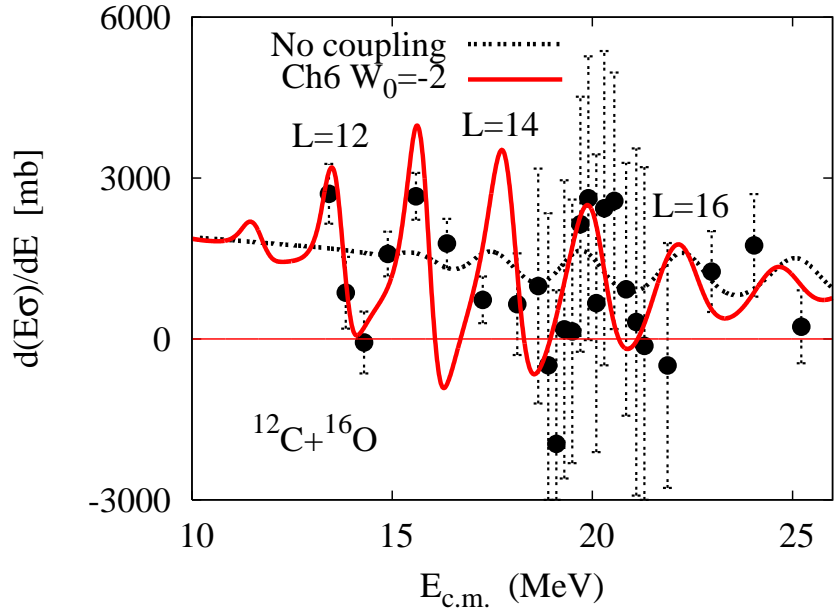}
\caption{\label{f1dc12o16} 
(Color online)
First derivative of some of the energy-weighted cross sections shown in 
Fig. \ref{flfc12o16}. Both calculations include an imaginary potential
with $W_0$ = -2 MeV. The peaks associated with $L$ = 12, 14 and 16 
in the Ch6 calculation are labeled.}
\end{figure}

The first derivative of the energy weighted cross sections is shown
in Fig. \ref{f1dc12o16}. By comparing the data to the coupled-channels 
calculation (the one with the weak absorption, $W_0$ = -2 MeV) it is 
possible to assign an angular momentum to each of the observed peaks.
Since the fusing systems is asymmetric, fusion occurs for both even 
and odd values of $L$.  Calculated barriers exist for all values of 
angular momenta in the range $L$ = 11-17, and the barriers for 
$L$ = 12, 14, and 16 are marked for clarity in the figure. 
Experimental barriers are clearly identified for $L$ = 12, 13, and 
15 but the barrier for $L$ = 14 is apparently missing. 

\begin{figure}
\includegraphics[width = 12cm]{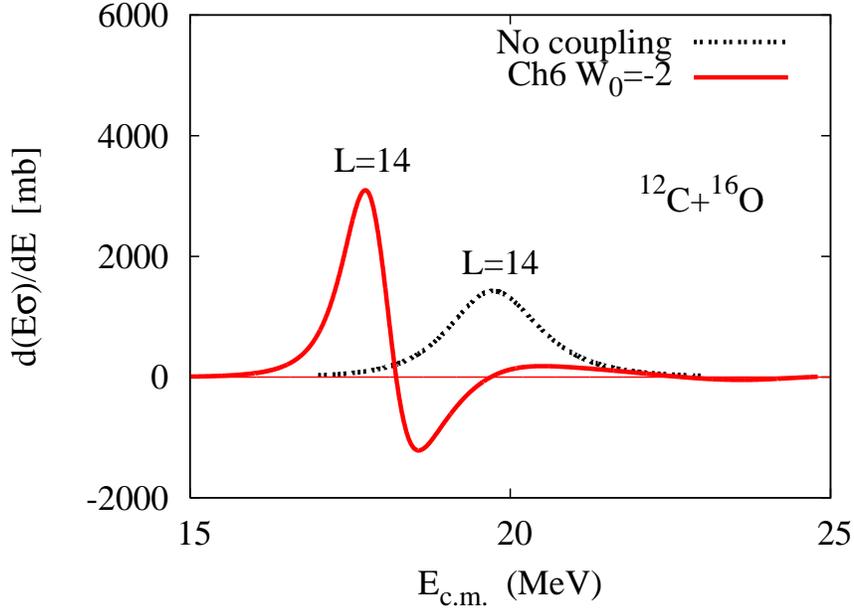}
\caption{\label{f1dc12o16L14} 
(Color online)
First derivative of the energy-weighted cross section 
obtained from the $L$ = 14 partial wave. The peak of the 
coupled-channels calculation Ch6 is lowered by about 2 MeV 
compared to the peak obtained in the no-coupling limit.}
\end{figure}

It is remarkable that the Ch6 coupled-channels calculation shown
in Fig. \ref{f1dc12o16} reproduces the peak structures of the 
data so well, both in position and in absolute magnitude. 
The no-coupling limit, on the other hand, produces very modest
peaks, and their positions are shifted compared to the peaks 
of the coupled-channels calculation. 
This is seen more clearly in Fig. \ref{f1dc12o16L14}, where the 
first derivative of the energy weighted contribution to the cross 
section from the orbital angular momentum $L$=14 is shown.
It is seen that the peak of the coupled-channels calculation 
is lowered by about 2 Mev compared to the no-coupling limit.
This implies that the $L$=14 peak obtained in the no-coupling 
limit is located in Fig. \ref{f1dc12o16} near the $L$=15 peak
of the coupled-channels calculation.

In summary, structures due to individual centrifugal barriers are 
clearly observed in the fusion data for all three combinations of
$^{12}$C and $^{16}$O nuclei.
The coupled-channels calculations reproduce most of the observed 
structures fairly well, provided the M3Y+repulsion entrance channel 
potential is relatively shallow and the imaginary potential 
is relatively weak and short ranged.
Using instead a conventional Woods-Saxon potential, the structures 
at low energies become suppressed, whereas the structures at higher
energies become much stronger (see Fig. 5B). 
The results for $^{12}$C+$^{16}$O demonstrate that coupled-channels
effects can be very large and shift the location of the effective
centrifugal barrier to lower energies.
The possibility of observing similar structures in the fusion of  
heavier systems is discussed in the next section.   

\section{Applications to heavier systems}

The search for structures in the high-energy fusion of heavier systems is 
difficult. The reason is that structures associated with individual angular 
momentum barriers can only be seen at high angular momenta in heavy systems, 
according to Eqs.  (\ref{condi1},\ref{condi2}). 
Since many reaction channels are expected to open up at high angular momenta 
and high energies in  heavy systems, the effect of couplings to these channels 
may smear out the peak structures. 
It is of great interest to pursue the search for such structures 
because they can provide valuable information (if they exist) about 
the ion-ion potential and put constraints on coupled-channels calculations.
On the other hand, the disappearance of the structures may indicate where
the coupling to many open reaction channels sets in. 

An experimental search was performed by Gary and Volant \cite{gary}
who investigated the fusion of $^{24}$Mg+$^{24}$Mg, $^{28}$Si+$^{28}$Si 
and similar systems. The amplitudes of the oscillations that were 
observed are modest compared to the oscillations that are seen in 
the fusion of the $^{12}$C and $^{16}$O systems. They are sometimes 
comparable to the experimental uncertainties which makes it difficult 
to judge whether the structures exist or not. It was concluded 
\cite{gary} that $^{12}$C+$^{24}$Mg and $^{28}$Si+$^{28}$Si are 
the only systems that exhibit an oscillatory behavior in the fusion 
data at high energy. Although these findings are disappointing, it is 
useful to analyze the data the same way it was done in the 
previous section, namely, in terms of the first derivative of the 
energy weighted cross section. It is of particular interest to see
whether the observed structures can be reproduced by coupled-channels
calculations.  As an example of a system that exhibits some structures, 
the fusion of the symmetric system $^{28}$Si+$^{28}$Si will be 
discussed in the following.

\begin{figure}
\includegraphics[width = 10cm]{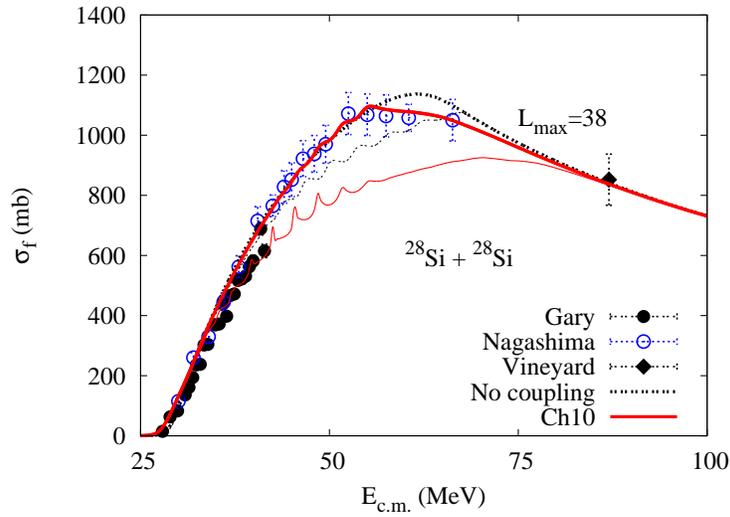}
\caption{\label{flfsi28he} 
(Color online)
Measured cross sections for the fusion of $^{28}$Si+$^{28}$Si 
\cite{gary,moto,vine} are compared to coupled-channels calculations 
(Ch10, solid curves) and to calculations without couplings
(dashed curves). The thin curves are based on a weak imaginary 
potential, with $a_w$=0.2 fm, $W_0$=-2 MeV, whereas the thick curves
use a stronger imaginary potential, with $a_w$=0.5 fm, $W_0$=-10 MeV. 
All calculations assume a maximum angular momentum of $L_{max}$=38.}
\end{figure}

It is difficult to calibrate the M3Y+repulsion interaction
to the $^{28}$Si+$^{28}$Si data by Gary and Volant \cite{gary} 
because they cover a relatively small range of energies. Fortunately, 
there is another data set by Nagashima et al. \cite{moto} which covers
a much broader range of energies, and the high energy data by Vineyard 
et al. \cite{vine} are also very valuable because they determine a 
limiting or critical angular momentum for fusion, which is about 
$L_c$ = 38$\hbar$. 
The three data sets are shown in Fig. \ref{flfsi28he}.
Also shown in this figure are coupled-channels calculations and
calculations performed in the no-coupling limit.  All calculations 
are based on an M3Y+repulsion potential, which is produced by $^{28}$Si 
densities of radius $R$ = 3.17 fm  and diffuseness $a$ = 0.48 fm. 
The diffuseness associated with the repulsive part of the interaction 
(see Ref. \cite{misi2} for details) was adjusted to $a_r$ = 0.378 fm
so that the data by Nagashima et al. were reproduced (see below.)

The structure input to the calculations is shown in Table I. 
The calculations include the excitation of the $2^+$, the $3^-$, and 
an effective two-phonon quadrupole states in each nucleus.
The two-phonon state was constructed following the procedure described 
in Ref. \cite{alge} from the information given in Table I about the 
$0_2^+$ and $4^+$ states. Unfortunately, there is no information
available about the $2_2^+$ state so it is ignored.
The mutual $(2^+,2^+)$, $(2^+,3^-)$ 
and $(3^-,2^+)$ excitations in projectile and target were also included, 
whereas the mutual excitation of the $3^-$ states was ignored because the 
excitation energy is so high. In addition to the elastic channel, that
gives a total of (1+3+3+3) 10 channels and the calculation is referred 
to as the Ch10 calculation.  

The nucleus $^{28}$Si is deformed with an oblate quadrupole shape.
The measured quadrupole moment of the $2^+$ state, $Q_2$ = 0.16(3) b 
\cite{stone}, implies a deformation parameter which is 
$\beta_2$ = -0.40(8). 
This is consistent with the measured $B(E2)$ value given in Table I. 
The quadrupole deformation was considered explicitly in the 
coupled-channels calculations by including the diagonal matrix 
element of the quadrupole interaction in the excited $2^+$ state 
(see Ref. \cite{alge} for details.)

It is difficult to make a good calibration of the M3Y+repulsion interaction 
without access to any data at subbarrier energies. The cross sections
at high energies are sensitive to other reaction mechanisms that are not
considered explicitly in coupled-channels calculations, so there is some 
ambiguity in the calibration of the ion-ion potential and the imaginary 
potential.  The thick solid curve in Fig. \ref{flfsi28he} is a compromise 
which does a fairly good job in reproducing the data of Nagashima et al. 
\cite{moto}. It is based on a relatively strong imaginary potential, 
$W_0$ = -10 MeV and $a_w$ =0.5 fm. 
Moreover, a maximum angular momentum for fusion, $L_{max}$ = 38, 
was imposed on the calculations in order to be consistent with the 
high energy data of Vineyard et al. \cite{vine}. 
The thin solid curve is the result of a similar 
calculation that is based on a weaker and short-ranged imaginary 
potential, $W_0$ = -2 MeV and $a_w$ = 0.2 fm.
It does not account for Nagashima's data at high energies but it is 
in fair agreement with the data by Gary and Volant.

\begin{figure}
\includegraphics[width = 10cm]{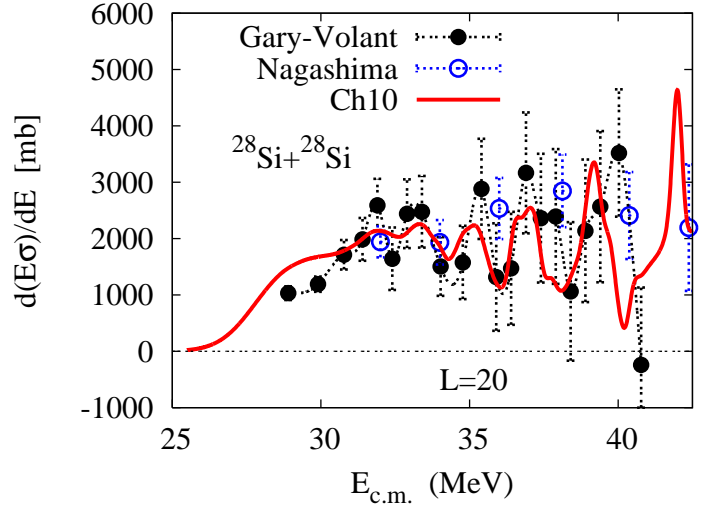}
\caption{\label{f1dsi28} 
(Color online)
First derivative of the energy-weighted cross sections shown in 
Fig. \ref{flfsi28he}. The coupled-channels calculation Ch10 is 
based on the weak imaginary potential, with $a_w$=0.2 fm, $W_0$=-2 MeV. 
The calculated peak for $L$=20 near 35 MeV is indicated.}
\end{figure}

The first derivative of the energy-weighted cross section for the fusion
of $^{28}$Si+$^{28}$Si is shown in Fig. \ref{f1dsi28}. The data by Gary 
and Volant \cite{gary} are connected by the dashed curve for clarity.
The data by Nagashima et al. \cite{moto} are also shown. They are 
consistent with the data by Gary and Volant but the energy steps between 
the data points are too large to reveal any structures in the data.

The solid curve in Fig. \ref{f1dsi28} is derived from the coupled-channels
calculation shown in Fig. \ref{flfsi28he} with  the weak absorption. 
It shows a lot of structure and it is remarkable how well the peaks of 
the calculation correlate with the structures observed in the data. 
The good agreement in this respect allows one to assign with some
confidence an angular momentum to each of the peaks. The peak near 35 MeV
is caused by the $L$ = 20 centrifugal barrier. 



Based on the results discussed here it is suggested that some of the 
measurements by Gary and Volant \cite{gary} should be repeated with higher 
precision and with sufficiently small energy steps so that the structures 
associated with the individual centrifugal barriers can better be resolved.  
In order to be able to calibrate the ion-ion potential, it would be very 
useful not only to perform the measurements at energies above the Coulomb 
barrier, where the structures due to the individual centrifugal barriers 
may exist, but also at energies far below the Coulomb barrier, where the 
sensitivity to the ion-ion potential for overlapping nuclei also shows 
up \cite{misi2}. 

The analysis of the $^{28}$Si fusion data shows that it is necessary to 
employ an imaginary potential of varying strength, depending on which 
data set is analyzed. The high-energy data require a strong imaginary
potential, combined with a maximum angular momentum for fusion, whereas 
the low energy data can be explained with a weak imaginary potential. 
It would be useful in future work to develop an energy 
dependent imaginary potential.

\section{Conclusions}

The structures that have been observed a long time ago in the 
high-energy data for several light-ion systems can be explained as 
being caused by the penetration of successive centrifugal barriers that 
are well separated in energy. This mechanism is clearly seen in 
Hill-Wheeler's expression for the fusion cross section, and it is best 
illustrated by plotting the first derivative of the energy weighted cross 
section. The locations of the peaks in such a plot show the energies of 
the centrifugal barriers that causes the structures, whereas the width 
of the peaks can be associated with the quantum mechanical penetration 
of the centrifugal barrier.  
The analytic expression for the fusion cross section derived by Wong,
on the other hand, does not reveal the energy location of any individual
centrifugal barriers, except the location of the s-wave barrier.
This is not surprising because the derivation of Wong's
formula assumes that sequential barriers are so close
in energy that the discrete sum over the orbital angular momentum be 
replaced by a smooth integration. 

Some of the fusion data were analyzed by coupled-channels calculations
that were based on the M3Y+repulsion potential. The calculations showed
that the strength and the location of the peaks observed in the first 
derivative of the energy-weighted cross section are very sensitive to 
coupled-channels effects. 
This implies that the extracted barriers are effective barriers and not
the real barriers of the centrifugal potential in the entrance channel.

The nuclear potentials that were used in the coupled-channels calculations 
were adjusted in each case to optimize the fit to the data, and when this 
was achieved, it turned out that the calculations reproduced fairly well 
the location of the peak structures that are observed in the data. 
The good agreement allows one to assign an orbital angular momentum to 
most of the effective centrifugal barriers that have been extracted from 
the experiments.

The results of the data analysis suggest that the structures observed in
the fusion data at energies above the Coulomb barrier are best explained 
by coupled-channels calculations that are based on a shallow potential in
the entrance channel. Thus there appears to be some consistency in the 
description of the structures at energies above the Coulomb barrier and 
the hindrance of fusion, which is observed at energies far below the 
Coulomb barrier. Both phenomena are sensitive to the ion-ion potential for 
overlapping nuclei, and both are best described by a shallow potential 
in the entrance channel.

The amplitude of the structures observed in the fusion data seems to 
diminish in heavier systems. This is unfortunate because the structures 
reveal valuable information about the ion-ion potential, and they provide 
an excellent test of coupled-channels calculations. It is very encouraging 
to see, however, that structures do exist in the fusion of a system as 
heavy as $^{28}$Si+$^{28}$Si. It is therefore suggested that a new 
experimental search for structures in high-energy fusion cross sections
be pursued, preferably with higher precision than in the past, and with 
energy steps that are sufficiently small to resolve the structures 
associated with the individual barriers.


{\bf Acknowledgments}. 
This work was supported by the U.S. Department of Energy, 
Office of Nuclear Physics, contract no. DE-AC02-06CH11357.

\begin{table}
\caption{The properties of the low-lying states in $^{28}$Si are 
from Ref. \cite{ENDSF}. The $0_2^+$ and $4^+$ states are lumped 
together into one effective two-phonon quadrupole state as described 
in Ref. \cite{alge}. Unfortunately, no information about the $2_2^+$
state is available. The Coulomb and nuclear deformation parameters are 
assumed to be the same.}
\begin{tabular} {|c|c|c|c|c|}
\colrule
  $I^\pi$ &  E$_x$ (MeV) & Transition & B(E$\lambda$) (W.u.) &
$\beta_\lambda^C$ \\ 
\colrule
 2$_1^+$  & 1.779 & $2_1^+$ - $0_1^+$ & 13.2(3)  & -0.411 \\ 
 0$_2^+$    & 4.980 & $0_2^+$ - $2_1^+$ &  8.6(16) &        \\ 
 4$^+$    & 4.618 & $4^+$ - $2_1$   & 13.8(13) &        \\ 
 Eff 2PH  & 4.689 &   2PH - $2_1^+$ &  8.8     & -0.238 \\ 
 3$^-$    & 6.879 & $3^-$ - $0_1^+$ & 13.9(24) &  0.416(35) \\ 
\colrule
\end{tabular}
\end{table}


\end{document}